# Atomic-scale mechanisms of defect- and light-induced oxidation and degradation of InSe

Andrey A. Kistanov,[a, b] Yongqing Cai,[*, b] Kun Zhou,[*, a] Sergey V. Dmitriev[c, d] and Yong-Wei Zhang[*b]

Layered indium selenide (InSe), a new two-dimensional (2D) material with a hexagonal structure and semiconducting characteristic, is gaining increasing attention owing to its intriguing electronic properties. Here, by using first-principles calculations, we reveal that perfect InSe possesses a high chemical stability against oxidation, superior to $MoS_2$. However, the presence of intrinsic Se vacancy ($V_{Se}$) and light illumination can markedly affect the surface activity. In particular, the excess electrons associated with the exposed In atoms at the $V_{Se}$ site under illumination are able to remarkably reduce the dissociation barrier of $O_2$ to ~0.2 eV. Moreover, at ambient conditions, the splitting of $O_2$ enables the formation of substitutional (apical) oxygen atomic species, which further cause the trapping and subsequent rapid splitting of $H_2O$ molecules and ultimately the formation of hydroxyl groups. Our findings uncover the causes and underlying mechanisms of InSe surface degradation via the defect-photo promoted oxidations. Such results will be beneficial in developing strategies for the storage of InSe material and its applications for surface passivation with boron nitride, graphene or In-based oxide layers.

## Introduction

Oxidation plays a paramount role in affecting the integrity, property and performance of materials. For instance, surface oxidation can make chemical catalysis inert,[1] while it can also be used to modulate material properties, such as band gap opening of graphene.[2] In two-dimensional (2D) materials, oxidation may lead to the structural degradation, thus decreasing their performance and impeding their applications.[3] For atomically thin 2D materials, two factors, light illumination[4,5] and atomic defects,[6,7] are particularly eminent in affecting their oxidation in comparison with three-dimensional (3D) counterparts. The photo-induced effect tends to be pronounced in semiconducting layered materials due to their broadband photoresponse inherent in their tunable and quantum-confined electronic states.[8] The weak electronic screening in atomically thin materials often triggers a strong light–matter interaction.[9] In addition, 2D materials are also prone to forming atomic defects owing to their ultrathin sheets.[10,11] Due to the break-up of lattice periodicity, redistribution of the electron density around defects core often occurs, rendering 2D sheets very active and sensitive to the environment.[6,12,13] Phosphorene is a well-known example which shows a rapid oxidation due to chemical adsorption of $O_2$ molecules,[14] photo-oxidation,[4,5] and defect-assisted oxidation.[15]

Indium selenide (InSe), a recently emerging layered metal monochalcogenide III-VI compound with each InSe layer composed of covalently bonded Se–In–In–Se atomic planes, has attracted great attention,[16,17] owing to its intriguing electronic properties and dramatically different behaviors compared with other 2D materials.[18,19] For instance, in sharp contrast to $MoS_2$ with the well-known direct (monolayer)-indirect (multilayer) transition, InSe has an opposite thickness-dependent behavior with an indirect band gap for monolayer while a direct band gap for multilayer sheets when exceeding a critical thickness.[20,21] The thickness-dependent direct band gap of thick InSe sheets allows a broad excitonic emission.[22] These unique properties trigger many studies on the growth,[23-25] exfoliation,[26] and applications of InSe, for example, in optoelectronics,[27-30] sensors, and photovoltaics.[19,28,31] However, the performance of InSe-based transistors at ambient conditions[16] or under external fields[32] was found to be unsatisfying due to the degradation of their performance.[33,34] Its lone-pair states of Se atoms at the top of the valence band of InSe induce high sensitivity to external adsorbates.[35] The situation was improved for the passivated InSe sheet, particularly, the electron mobility of the boron-nitride/graphene passivated InSe sheet was found to exceed $10^3$ $cm^2$ $V^{-1}s^{-1}$ at room temperature.[36] Previous experiments revealed that thinner InSe films tend to suffer from a more rapid degradation, largely in the form of oxidation, compared with bulk InSe.[16] Therefore, understanding the degradation mechanisms of InSe that are predominantly involved with external adsorbates, such as $O_2$ and $H_2O$ at ambient conditions, is thus critically important for its practical applications. However, such understandings are still lacking. In particular, the knowledge of structure degradation arising from defect-environment-light coupling remains largely unknown.

In this work, by using *ab initio* electronic structure calculations and *ab initio* molecular dynamics (AIMD) simulations, we explore the mechanisms of InSe oxidation. We consider elementary steps of oxidation of monolayer InSe via $O_2$ and $H_2O$ molecules at ambient conditions, by taking into account the roles of defects. Our calculations reveal that perfect monolayer InSe has a relatively low oxygen affinity and thus is less likely to be oxidized. However, the presence of Se vacancies and exposure to light can greatly activate the InSe surface and facilitate the chemical dissociation of $O_2$. The partially oxidized InSe can make the surface hydrophilic and favor the adsorption of water, which may explain the degraded performance of unprotected samples at ambient conditions.

**Computational details**

Our theoretical calculations are performed within the framework of spin-polarized density functional theory (DFT) using Vienna *ab initio* simulation package (VASP).[37] The Perdew−Burke−Ernzerhof exchange-correlation functional under the generalized gradient approximation together with the van der Waals corrected functional with the Becke88 optimization is implemented for the geometry optimization. To achieve a high accuracy in describing electronic interactions, the Heyd–Scuseria–Ernzerhof (HSE06)[38] functional is used in the band structure calculations. All the structures are fully relaxed until the atomic forces are smaller than 0.01 eV/Å. The relaxed lattice constant of monolayer InSe is $a = b = 4.102$ Å. To model the defective and gas-adsorbed InSe, as well as for AIMD simulations, we adopt a 4×4×1 supercell and the corresponding k-mesh is 10×10×1. The plane-wave cut-off energy is 400 eV. A vacuum space with a thickness of 20 Å is introduced along the out-of-plane direction. Negatively charged $O_2$ ($O_2^-$) molecule is introduced by adding one electron to the considered system. The Bader analysis[39] is used for the charge transfer calculations. The reaction barriers are calculated using the climbing image nudged elastic band method (CI-NEB).[40] AIMD simulations are performed at room temperature of 300 K using the Nose-Hoover method with a time step of 1.0 fs.

**Results and discussion**

**A. Band alignment and indirect-direct gap crossover of few-layer InSe.** Oxidation of a 2D material at ambient conditions can involve three chemical steps (Figure 1a): adsorption of $O_2$ molecules, dissociation of $O_2$ molecules, and interaction of $H_2O$ molecules with the anchored oxygen species. Before examining the light- and defect-assisted chemical dissociation of $O_2$ molecules on InSe, we first study the electronic properties of pristine InSe and the mechanism of light-induced electron-hole pairs in InSe. The band structure of monolayer InSe is shown in Figure 1b indicating an indirect band gap ($E_g$) of 2.12 eV (HSE value). The striking feature of layered InSe is that for monolayer (1L) InSe and few-layer ($n$L) InSe with its layer number ($n$) below a critical value of ($n_c$), the conduction band minimum (CBM) is always located at Γ point, while the position of the valence band maximum (VBM) sits at the Λ point between the Γ and K points (refer to the band structures of 1L-4L InSe in Figure S1 in Supporting Information). Above the $n_c$ value (the exact value of $n_c$ is still under debate: the experimental value of the critical thickness of 6 nm with $n_c$ equal to 7 according to Ref 41 and theoretical result of 28 according to Ref. 42), the few-layer and bulk InSe become a direct $E_g$. Figure 1c schematically shows this evolution of the band edges of few-layer InSe. The Λ point gradually shifts towards the Γ point with an increase the number of layers and coincides with the Γ point upon $n > n_c$. The underlying reason for this phenomenon may originate from the interlayer coupling of the atomic lone-pair electrons between Se atoms in neighboring layers. This indirect ($n < n_c$) to direct ($n > n_c$) transition behavior is in sharp contrast to transitional metal dichalcogenides (TMDs) where only monolayer TMDs sheet has a direct $E_g$ and few-layer TMDs have an indirect $E_g$, suggesting different interlayer coupling mechanisms between InSe and TMDs.

The crossover to the direct gap for InSe sheets above the critical thickness $n_c$ suggests that efficient light adsorption is favored in multilayer InSe with more than $n_c$ layers. However, for the indirect-gap InSe with $n < n_c$ layers, as the Λ point of the VBM is very close to the Γ point, light adsorption can still occur through phonon-involved excitations to remedy the momentum mismatch. Such a phonon-assisted process is schematically shown in the inset of Figure 1c. It was shown that phonon-coupled phenomena can account for the luminescence of quantum wells and epitaxial layers.[42,44] In particular, for few-layer indirect-gap $MoS_2$, $MoSe_2$, and $WSe_2$, such a phonon-assisted process is able to lead to appreciable photoluminescence.[45-47] However, for few-layer $MoS_2$, the CBM lies around the middle of the Γ-K point and the VBM at Γ, implying a significant momentum mismatch, and thus the number of phonons eligible for such a process is quite limited. Herein, in monolayer and few-layer (n < $n_c$) InSe, the indirect-gap Λ point in the VBM is in the proximity of Γ, and thus the momentum mismatch is quite small. This allows a large number of long-wavelength flexural phonons to promote the light adsorption and trigger excitons. Indeed, significant excitation and strong light adsorption were observed in monolayer and few-layer InSe.[9,20]

Figure 1d shows the thickness-dependent position of the CBM and VBM for 1L-5L InSe aligned relative to the vacuum level ($E_{vac}$). The $E_g$ decreases from 2.13 eV for monolayer InSe to 1.08 eV for 5L InSe with an upward (downward) shift of VBM (CBM). Our predicted wide tunable $E_g$ may account for the observed broadband photoresponse in layered InSe.[48,49] In principle, upon light illumination, the photo-excited electrons can be transferred to the adsorbed $O_2$ molecules, thus affecting their adsorption and dissociation behaviors (refer to the schematic plot in Figure 1a). The effect of the photooxidation of InSe will be discussed shortly.

**B. Localized states of Se vacancies and molecular adsorption.** Besides the light adsorption, intrinsic defects, like the atomic vacancies, can also influence the electronic properties and act as active sites for molecule adsorption. Next, we examine the most possible intrinsic defects, that is, the Se vacancies ($V_{Se}$) in InSe. Figure 2a shows the density of states (DOS) of perfect monolayer

InSe together with those of the mono-vacancy (MV), and di-vacancy (DV) in monolayer InSe. To highlight the $V_{Se}$-induced effects, all the DOS curves are plotted with the VBM of host InSe shifting to zero. Each loss of a Se atom in the InSe sheet creates three dangling bonds associated with the unpaired orbitals of the three exposed In atoms. The unpaired orbitals in each MV $V_{Se}$ evolve into the two defective states in the DOS: a lower-lying singly "$A_1$" (occupied) state around 0.4 eV above the valence band and a doubly degenerate higher-lying "E" (unoccupied) state nearly coinciding with the edge of the conduction band. The two levels are named according to their symmetries in the irreducible representation of the local symmetry ($C_{3v}$) of the MV $V_{Se}$. The spatial distribution of both states is highly localized and plotted in the bottom panel of Figure 2a. It can be seen that both $A_1$ and E states are distributed across the top and bottom surfaces with significant weight on the In atoms while only slight components from the neighboring Se atoms. Note that the $A_1$ (E) is a bonding (antibonding) state of the middle In-In bond. With the removal of another Se atom neighboring a pre-formed $V_{Se}$, a DV $V_{Se}$ is thus formed, with more defective states than the MV case, reflected by the enhancement and broadening of peaks in the DOS plot in Figure 2a. The singly occupied "$A_1$" level of an MV $V_{Se}$ evolves into two slightly split levels in the DV case whereas the degenerate pair of "E" level splits considerably into four levels, originated from the relatively strong coupling of the dangling states of the two $V_{Se}$. The evolution of the defective states at Γ from the MV to DV-containing InSe is schematically shown in Figure 2b. Exact alignment and positions of these defective levels are more clearly reflected from the comparison of the band structures of perfect, MV- and DV-containing InSe as plotted in Figure 3a. It should be noted that to compare the band gap of perfect InSe with those of InSe with low defect concentrations, we calculate the band gaps of defective InSe by measuring the VBM and CBM related to the host InSe with the defective states being excluded. Our calculations show that the band gap increases with the defect concentration. More specifically, perfect InSe has the smallest band gap of ~1.4 eV. With the increase of the vacancy concentration from 2 to 3.13 and further to 6.26 %, the band gap size increases from 1.52 to 1.55 and further to 1.68 eV, respectively. To facilitate the comparison of the defective states, the band structures of the different systems in Figure 3a are all adjusted to align the VBM of host InSe. The presence of MV and DV $V_{Se}$ increases the $E_g$ of InSe by 0.16 and 0.33 eV, respectively, which can explain the experimental blue shift of the photoluminescence peak after thermal annealing (~35 meV at 175 °C).[32] It is expected that these defective states act as trapping and scattering centers for conducting carriers and decrease their electronic mobility.

The presence of defects on the surface of a material can greatly modify its physical properties in terms of energetics, kinetics and charge transfer during gas adsorption.[6,12,13] We examine the effect of $V_{Se}$ on the adsorption energy ($E_a$), which is calculated as $E_{InSe + mol} - E_{mol} - E_{InSe}$, where $E_{InSe+mol}$, $E_{mol}$, and $E_{InSe}$ are the total energies of molecule-adsorbed system, gas molecule, and InSe, respectively (compiled in Table S1 in Supporting Information). Our charge transfer analysis shows that the $O_2$ and $H_2O$ molecules are molecular acceptors for both the perfect and $V_{Se}$-containing InSe (refer to the differential charge transfer plots in Figures 3b and S2a, b). Compared with the adsorption on perfect InSe (Table S1), the presence of MV $V_{Se}$ triggers a much larger effect on the $H_2O$ molecule than on the $O_2$ molecule: the $E_a$ changes from -0.17 eV (perfect InSe) to -0.41 eV ($V_{Se}$ site) for $H_2O$, while there is almost no change in the $E_a$ for the $O_2$ (~ -0.10 eV) molecule. Similarly, for the Se-deficient InSe, the $H_2O$ molecule has a significantly enhanced oxidization ability compared with the perfect surface as judged from the amount of charge transfer between the molecule and the surface. For the $H_2O$ molecule, the amount of charge transfer is more than four times larger at the $V_{Se}$ site (-0.09 e) than at the perfect surface (-0.02 e), while for the $O_2$ molecule, the charge transfer (around -0.03 e) is almost insensitive to the presence of $V_{Se}$.

The band structures and DOS of the $O_2$ and $H_2O$ adsorbed on the MV $V_{Se}$ site (refer to Figure S2 in Supporting Information for adsorption on the perfect InSe) are shown in Figures 3a and b, respectively. Upon being adsorbed around the $V_{Se}$ center, the $O_2$ takes a parallel geometry and the $H_2O$ adopts a tilted configuration with the two H-O bonds pointing away from the surface. The lowest unoccupied molecular orbital (LUMO) state ($2\pi^*$) of the $O_2$ molecule is slightly below the empty "E" defective level, suggesting that any captured electrons in this defective level can be easily transferred to the $O_2$ molecule. For the $H_2O$ molecule, there is no $H_2O$ related state within the band gap of InSe. However, the presence of $V_{Se}$ makes the highest occupied molecular orbital (HOMO) state ($1b_1$) of $H_2O$ significantly downward shifted by ~2.2 eV compared to the adsorption on the perfect InSe (refer to Figure S2b in Supporting Information). This may be the underlying reason for the abovementioned large differences in the $E_a$ and the charge transfer for adsorptions of $H_2O$ on the $V_{Se}$ site compared with perfect InSe.

**C. Chemical dissociation of $O_2$ molecules and the effect of $H_2O$ molecules.** We next consider the kinetics of the $O_2$ molecule dissociation on the InSe. The detailed reaction path and the corresponding energetic profile for the $O_2$ decomposition during the surface oxidation are shown in Figure 4. We find that the oxidation process is exothermic and the barrier is strongly dependent on the surface stoichiometry of InSe and the charging state of the adsorbed $O_2$ molecule. The calculated barrier ($E_b$) for the dissociation of $O_2$ on perfect InSe is found to be 1.21 eV (Figure 4b, blue line). This value is significantly higher than that of for the $O_2$ chemisorption on phosphorene with $E_b$=0.56 eV[5] and on $MoS_2$ with $E_b$=0.74 eV,[6] suggesting a higher chemical stability of InSe against oxidation than phosphorene and $MoS_2$.

Upon light excitation, the $O_2$ molecule becomes superoxide anions and the $E_b$ is significantly reduced to 0.84 eV for the $O_2^-$. According to the Arrhenius formula, the reaction rate ($v$) depends exponentially on $E_b$ as $v = v_s \cdot \exp(-E_b/kT)$, where $v_s$ is the characteristic frequency (normally around $10^{13}$ Hz), k is the Boltzmann constant and $T$ is the temperature. It is expected that the drop of $E_b$ from 1.21 to 0.84 eV implies a 7-order higher oxidation rate. On the contrary, the barrier for the $O_2^+$ molecule increases to 1.32 eV. As shown in Figure 4a, starting from the physisorbed initial state (IS), the bond length of the $O_2$ molecule gradually increases and finally breaks. In the final state (FS), the two oxygen atoms adopt bridging configurations by forming In-O-Se and $In_2$-O-Se structures separately (Note that the plotted atomic configurations are based on neutral $O_2$ calculation). The case of $O_2^-$ undergoes similar structural transformations (refer to Figure S3 in Supporting Information). It should be noted that upon light illumination, the hole-doped monolayer InSe is not simulated in our calculations. In reality, the generated hole population within the InSe sheet and the superoxide anions at the surface may induce an electric field near the InSe surface region, which can make the barrier even smaller. Therefore, our results may explain the poor structural stability and unsatisfying mobility of the unpassivated InSe in the MOS architecture where the vertical field was created through applying gating voltage.[16]

Surprisingly, with the introduction of a single $V_{Se}$ in the InSe sheet, as shown in Figures 4c, d, the $E_b$ is dramatically reduced to 0.24 eV. This ultralow barrier suggests that the $O_2$ molecule can easily dissociate at the $V_{Se}$ site even at a moderate temperature and the Se-deficient InSe becomes easily oxidized at ambient conditions. The mechanism of the promoted activity may originate from the following reasons. Firstly, the loss of Se atoms allows the exposure of In atoms to oxygen where is a direct transfer of unpaired excess electrons of In to the 2π* antibonding state of $O_2$ molecule. Second, by contacting the strong electropositive In atoms with the strong electronegative O atoms, the formation of In-O bonds in the final state (FS) is accompanied by a strong release of energy. The energy difference of the transition state (TS) and an initial state (IS) in the MV $V_{Se}$-containing InSe achieves up to ~3 eV per $O_2$ molecule (Figure 4d), which is nearly three times the energy release of ~1 eV in the case of perfect InSe (Figure 4b). Finally, the 4d state of the exposed In atoms at the vacancy site may also facilitate the spin triplet-singlet transition of the $O_2$ molecule, which can contribute to the low barrier of $O_2$ dissociation. With the coexistence of light and $V_{Se}$, the $E_b$ can be further reduced and become almost negligible for the photo-reduced $O_2$ molecule ($O_2^-$, red line in Figure 4d, undergoes similar structural transformations, refer to Figure S4 in Supporting Information), suggesting that under a proper light illumination or charge injection, the oxidation rate of InSe can be dramatically increased even at room temperature.

There are two consequences arising from the chemical dissociation of the $O_2$ molecule on the InSe surface: One is related to the electronic properties. Upon the formation of the In-O-In structures in the Se-deficient InSe sheet (the TS structure shown in Figure 4c), the defect state due to $V_{Se}$ is quenched (see the LDOS in Figure S2c in Supporting Information). This may affect the optical properties via suppressing non-radiative recombination at the vacancy site and increasing the quantum efficiency. The second one is related to those embedding O atoms in InSe. Under ambient conditions, the bridging O atoms with In-O-In or In-O-Se bonding configurations or apical O atoms in forming O-Se vertical groups in the sheet may affect the adsorption of the polar molecules, like $H_2O$, on the InSe surface.

Next, we perform AIMD calculations to further investigate the adsorption of the $H_2O$ molecule by focusing on the pre-adsorbed O species on InSe. We compare the kinetics of the $H_2O$ molecule on three different InSe surfaces with different stoichiometric and oxidizing conditions: i) the perfect InSe without $V_{Se}$ and O groups, ii) the partially oxidized InSe without $V_{Se}$, and iii) the partially oxidized InSe with $V_{Se}$. The surface models for the latter two cases correspond to the end states of the NEB calculations shown in Figure 4. We analyze the cumulative distance $d_a$ for the atom $a$ at the $i$th MD step along the path $l_a$, which can be calculated by

$$d_a(N_i) = \int_{N_0}^{N_i} dl_a. \tag{1}$$

This analysis allows for the differentiation of the characteristics of the kinetic motion of $H_2O$. The cumulative distances and the trajectories of the O atom and one of the H atoms in the $H_2O$ molecule for the three cases are shown in Figure 5. For perfect pristine InSe, the trajectory (inset of Figure 5a) of the $H_2O$ molecule clearly shows a random walking behavior, which is also supported by the simulated snapshots in Figure S5 in Supporting Information. For the two partially oxidized InSe surfaces, that is, the pre-oxidized stoichiometric InSe (Figure 5b) and pre-oxidized Se-deficient InSe surfaces (Figure 5c), the kinetic behavior of the $H_2O$ molecule becomes significantly different, depending on the type of O groups on the surface. Note that the $H_2O$ molecule is initially positioned around the O group (with a distance around 3.5 Å) with only van der Waals interaction in order to reduce the computational time in sampling the phase space for the interaction of $H_2O$ with the adsorbed O atoms. For the pre-oxidized InSe without $V_{Se}$ (Figure 5b), the $H_2O$ molecule initially interacts with the terminated O atom in the apical Se-O group normal to the surface. Subsequently, the $H_2O$ molecule moves around the O atom. Interestingly, one of the O-H bonds in the water molecule is torn apart at 2.3 ps (the bottom right inset of Figure 5b). This is accompanied by the formation of two H-O-Se groups, signifying a spontaneous dissociation of the water molecule around the terminated O atoms. This bond breaking in $H_2O$ can be clearly seen from non-overlapping cumulative distance curves of the H and O atoms shortly after the adsorption (see the vertical dashed line in Figure 5b). The trajectory plots also clearly record this splitting with the H atom (green line) moving in the proximity of the terminated O atom on the InSe surface, while the left O (red line) and H (not shown) atoms in the initial $H_2O$ molecule gradually diffuse away. In contrast to the $O_2$ molecule, direct dissociation of $H_2O$ at the Se vacancy is impossible at a moderate temperature due to a large energy barrier of ~2.9 eV (see Figure S7 in Supporting Information).

Our AIMD simulations clearly show that besides the atomic oxygen species, the hydroxyl groups are also present on the InSe surface at ambient conditions (Figure S6 in Supporting Information). Moreover, at ~7 ps, one of the Se atoms bonded with O and H atoms is lifted off the surface by the outward dragging of the H-O group. However, the stability of the InSe structure is maintained with no disintegration in subsequent MD simulation in a longer timescale (Figure S6a in Supporting Information). The In atomic layer, which is behind the next layer of Se atoms, is also kept intact during the simulation.

For chemically bound –O and –OH groups above perfect InSe, defective states close to the valence band of InSe are introduced (see Figure S7 in Supporting Information), serving as carrier trapping centers. The adsorptions also induce p-type doping in InSe with charge transfer of 0.27 and 0.17 $e$ from the InSe surface to O and OH, respectively, amounting to $1.16 \times 10^{12}$ and $0.73 \times 10^{12}$ $e$/cm$^2$ according to our current atomic model with the O and OH coverage of 6.25% (see Figure S8 in Supporting Information).

Figure 5c shows the motion of the $H_2O$ molecule above the pre-oxidized Se-deficient InSe. As aforementioned in Figure 4c, there are two oxygen atoms occupying different local environments with one forming the O-[In]$_3$ group and the other forming the In-O-Se group. The $H_2O$ molecule is initially placed around the In-O-Se group (see the snapshot at 0 ps in the top left inset of Figure 5c). This type of O atom seems to have a weak effect and the water molecule moves randomly until it meets with the second O atom in the O-[In]$_3$ group at ~3 ps (see the snapshot in the top right inset of Figure 5c). The $H_2O$ molecule is trapped by this O atom and rotates strictly around it during the simulation time up to 20 ps. The trajectory plots with predominant lower-lying H atom close to the embedding O group suggest that the interaction mainly involves weak hydrogen bonding. Our findings here suggest that the presence of the atomic O group on the InSe surface can dramatically change the adsorption behavior of the water molecule. Controlling the type of the O groups like the in-plane O-[In]$_3$ group can enhance the adsorption of H$_2$O and increase the hydrophilicity of the surface.

**Conclusions**

In summary, we examined several critical issues in the structural degradation of InSe due to oxygen and humidity at ambient conditions using first-principles calculations. The oxidation of monolayer InSe was explored by examining the roles of light illumination, oxygen, water and defects. We showed that pristine perfect InSe has a much lower oxygen affinity than MoS$_2$ and phosphorene. However, the presence of V$_{Se}$ and light excitation can significantly accelerate the oxidation by greatly decreasing the barrier through forming chemical oxygen species. These atomic O species, which are associated with strong polar O-In bonds, can quench the defective states of V$_{Se}$, and further act as the adsorption and trapping centers of H$_2$O molecule. Our AIMD results showed that the apical O atoms in the form of terminated Se-O bonds even allow the spontaneous water splitting and the formation of hydroxyl groups at room temperature. Based on our findings, we propose the following three strategies to suppress the oxidation of InSe: i) insulating InSe from O$_2$ molecules; ii) maintaining the InSe surface stoichiometry; iii) avoiding the exposure of InSe to light illumination.

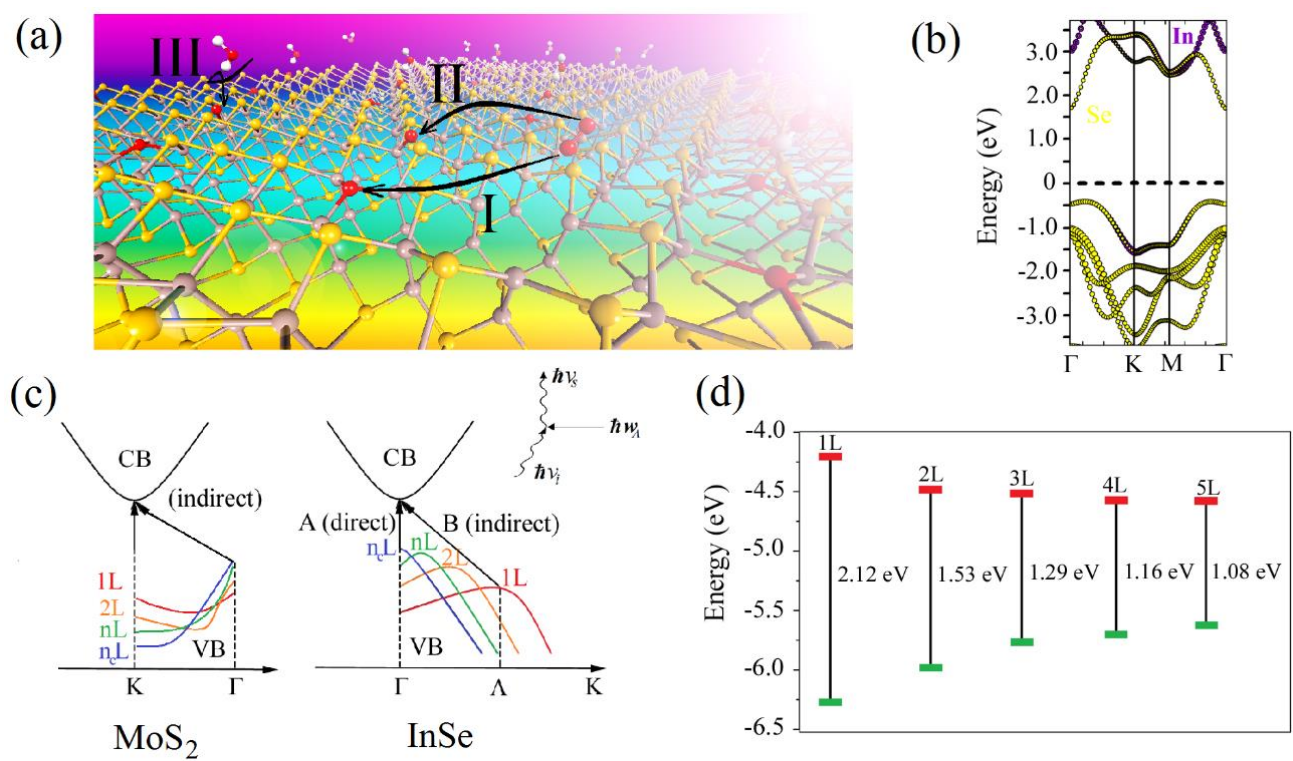

**Figure 1.** (a) Schematic for the photo facilitated oxidation of InSe. Step I: Formation of superoxide anions ($O_2 + h\nu \rightarrow O_2^- + h^+$); Step II: Chemical dissociation of $O_2^-$ into O atoms and oxidation of InSe; Step III: Adsorption of water molecules around the O atoms in the partially oxidized InSe. (b) Atomic projected HSE band structure for a perfect InSe sheet. (c) Schematic plot for the evolution of the valence band edges and the indirect-direct gap transition for InSe (right panel) and direct-indirect transition for MoS$_2$ (left panel) from 1L to $n$L, where $n$ is the number of layers. Two different light adsorptions with "A" and "B" excitation for direct-gap layers ($n > n_c$) and indirect-gap layers ($n < n_c$), respectively. The inset shows the schematic process of the phonon ($\hbar\omega_\Lambda$, momentum $\Lambda$)-assisted light adsorption and emission in few-layer indirect-gap InSe with the incident ($\hbar\nu_i$) and scattered ($\hbar\nu_s$) photons. (d) Thickness-dependent band alignment of the VBM and CBM of few-layer InSe with respect to the vacuum potential.

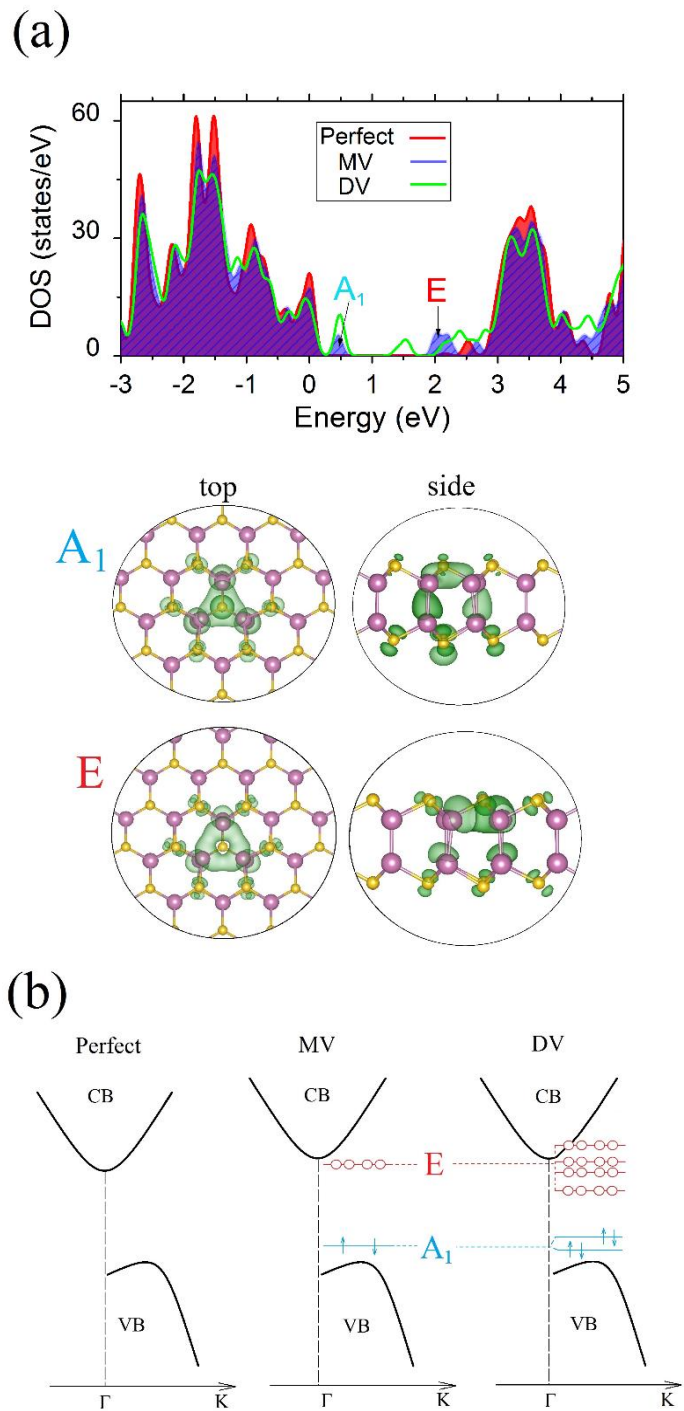

**Figure 2.** Localized defective states associated with the $V_{Se}$ in InSe. (a) DOS of perfect and defective InSe containing mono-(MV) and di-vacancy (DV) of Se atoms. All the energy levels are referenced to VBM of InSe. The $A_1$ and E correspond to the occupied and empty defective levels associated with MV $V_{Se}$, respectively. The inset shows the top and side views of the spatial distributions of the $A_1$ and E states around the MV $V_{Se}$. The In and Se atoms are colored in purple and yellow, respectively. (b) Relationship and evolution of the defective levels of MV and DV.

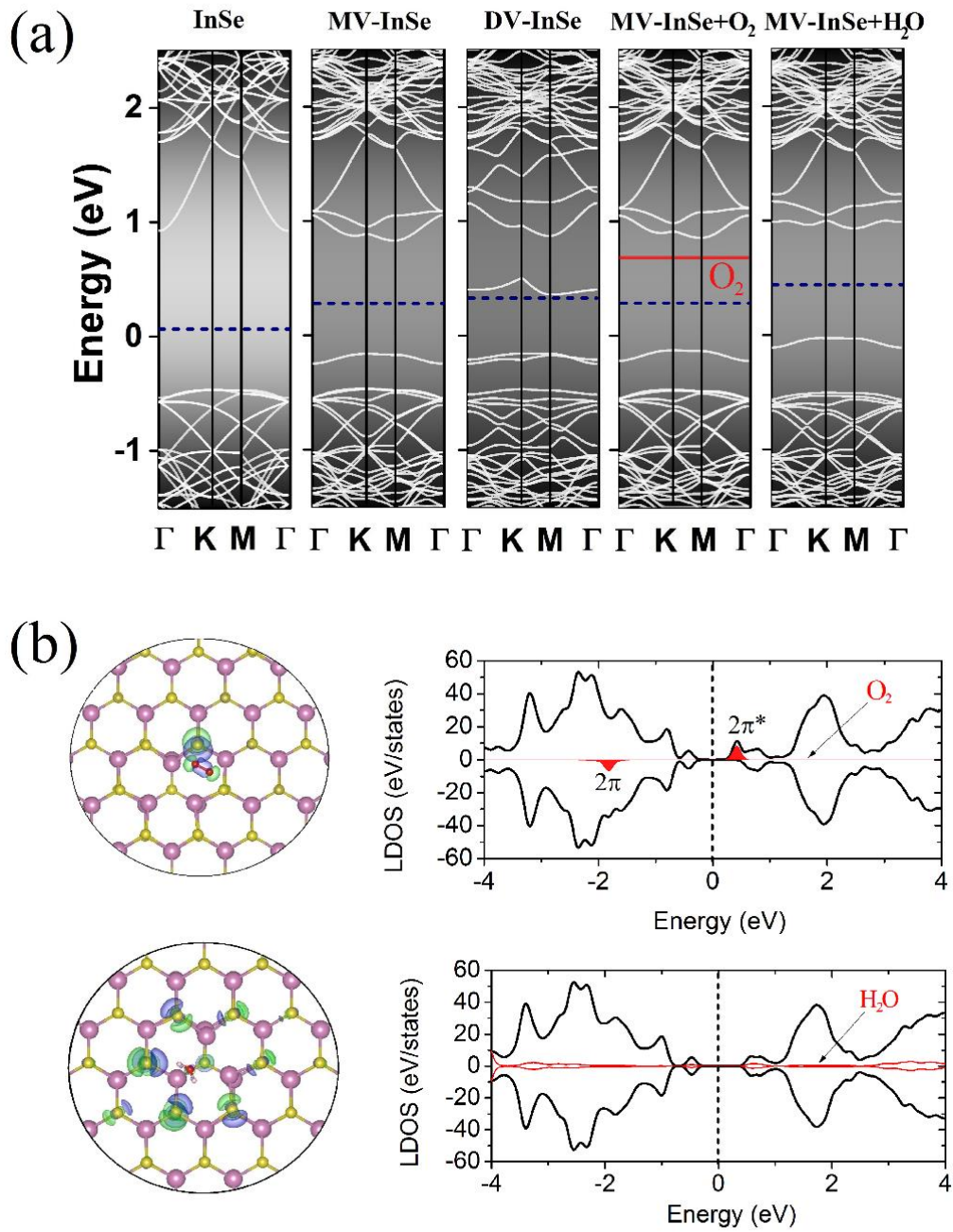

**Figure 3.** (a) Comparison of the band structures (GGA method) of perfect monolayer InSe, $V_{Se}$-containing InSe (MV and DV), and $O_2$/$H_2O$ physisorbed MV-InSe. Note that all the bands in different systems are shifted to align with VBM of the host InSe. (b) Isosurface plots of the differential charge density after $O_2$/$H_2O$ physisorption, where the green/blue color denotes depletion/accumulation of electrons (left panel) and DOS for $O_2$/$H_2O$ molecule adsorbed on the $V_{Se}$ site (right panel) with the Fermi level (dashed line) aligned at zero. States of $O_2$/$H_2O$ (total system) are denoted by the red (black) lines.

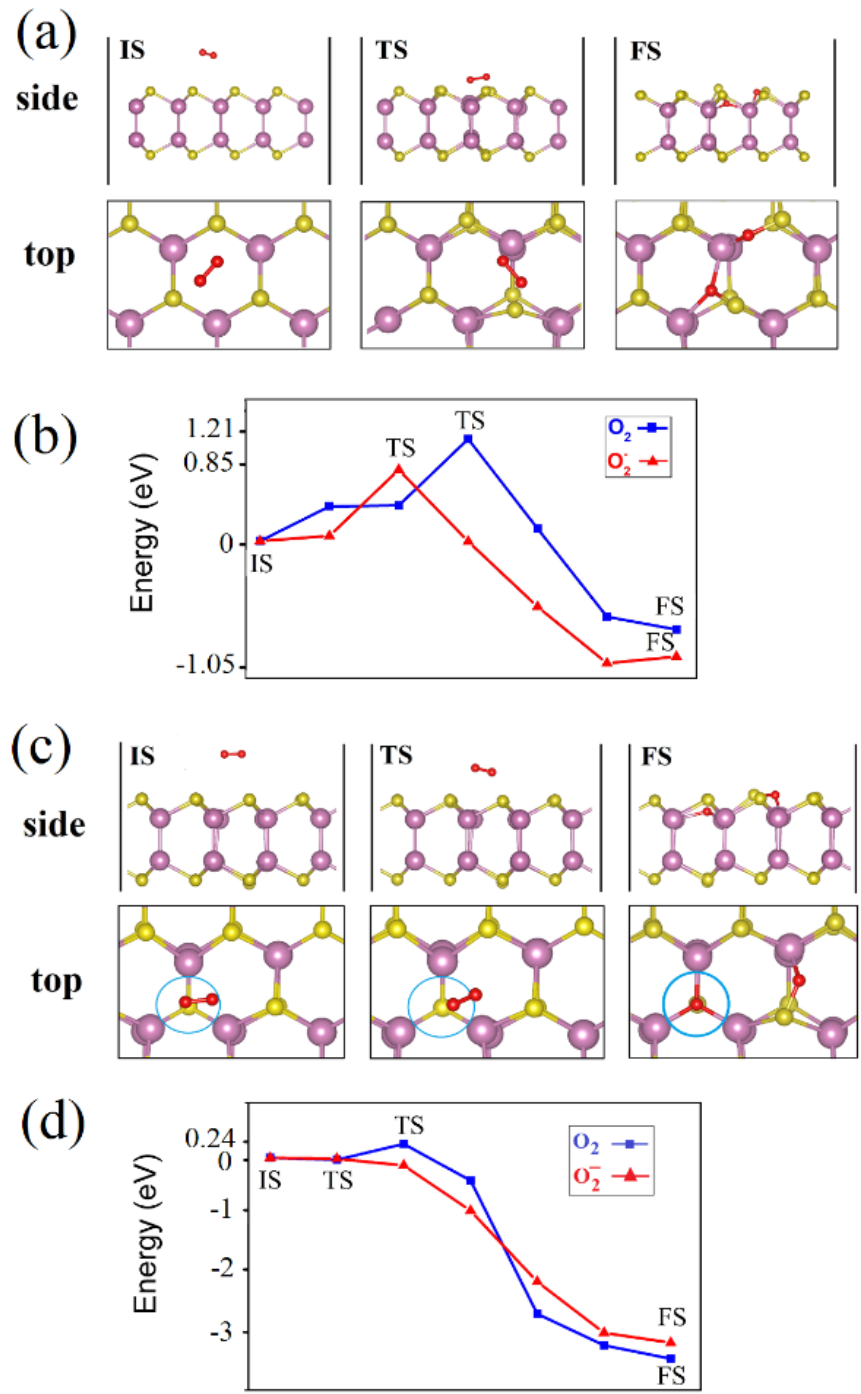

**Figure 4.** Chemical dissociation of the $O_2$ molecule on perfect (a, b) and MV-$V_{Se}$-containing (c, d) InSe. (a, c) Atomic configurations from the physisorbed to the chemisorbed state in the dissociation process of $O_2$. (b, d) Energetic profiles of the reaction pathway calculated by CI-NEB calculations. The IS, TS and FS represent the initial, transition, and final states, respectively. In, Se, and O atoms are colored in purple, yellow, and red, respectively. The position of the $V_{Se}$ is represented by the circle in (c).

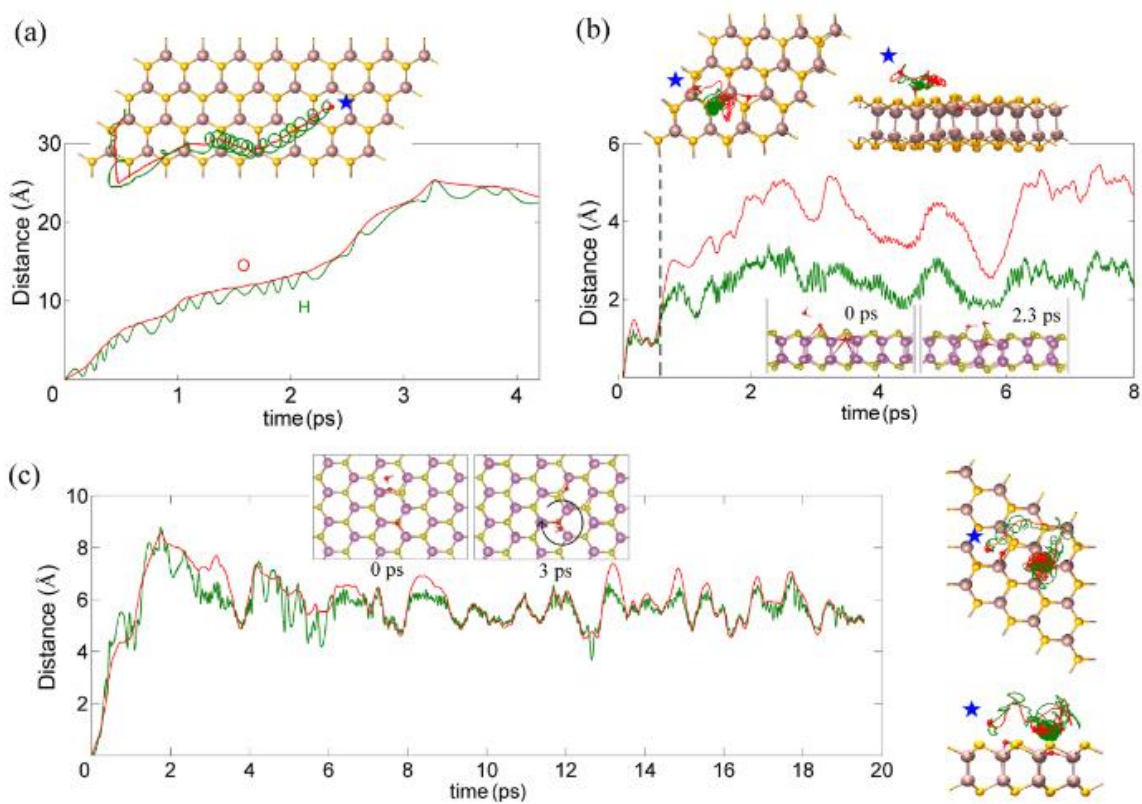

**Figure 5.** Kinetics of the H$_2$O molecule on the InSe surface by AIMD simulations at 300 K. Cumulative distance of the O atom (red curve) and one of the H atoms (green curve) in the H$_2$O molecule adsorbed on perfect InSe (a), partially oxidized InSe without V$_{Se}$ (b) and with V$_{Se}$ (c). The trajectories of the O (red curve) and H (green curve) atoms diffusing on the InSe surface are shown in the insets. The blue stars in the trajectory plots indicate the starting point of the H$_2$O molecule. AIMD snapshots in the insets of (b) and (c) show the splitting and the rotation of the H$_2$O molecule, respectively.